\documentclass[a4paper]{jpconf}

\usepackage{graphicx}
\usepackage{lmodern}
\usepackage{amsmath,amsfonts}

\begin{document}
\title{Transport coefficients from the Nambu-Jona-Lasinio model for $SU(3)_f$}

\author{R.~Marty$^1$, E.~Bratkovskaya$^1$, W.~Cassing$^2$, J.~Aichelin$^3$, H.~Berrehrah$^1$}

\address{$^1$ Frankfurt Institut for Advanced Studies and Institute for Theoretical Physics, Johann Wolfgang Goethe Universit\"at, Frankfurt am Main, Germany}
\address{$^2$ Institut f\"ur Theoretische Physik, Universit\"at Gie\ss{}en, Gie\ss{}en, Germany}
\address{$^3$ Subatech, UMR 6457, IN2P3/CNRS, Universit\'e de Nantes, \'Ecole des Mines de Nantes, Nantes, France}

\ead{marty@fias.uni-frankfurt.de}

\begin{abstract}
We calculate the shear $\eta(T)$ and bulk viscosities $\zeta(T)$ as well as the electric conductivity $\sigma_e(T)$ and heat conductivity $\kappa(T)$ within the Nambu-Jona-Lasinio (NJL) model for 3 flavors as a function of temperature as well as the entropy density $s(T)$, pressure $P(T)$ and speed of sound $c_s^2(T)$. We compare the results with other models such as the Polyakov-Nambu-Jona-Lasinio (PNJL) model and the dynamical quasiparticle model (DQPM) and confront these results with lattice QCD data whenever available. This work is based on Ref. \cite{Marty2013}.
\end{abstract}

\setlength{\abovedisplayskip}{2mm}
\setlength{\belowdisplayskip}{2mm}

\section{Introduction}

In order to study the expansion of the partonic plasma created in ultra-relativistic heavy-ion collisions \cite{Marty2012}, it is very helpful to calculate and compare thermodynamic properties as well as transport coefficients in equilibrium as a function of the temperature $T$. These results are not easily obtained from lattice QCD calculations and one has to consider suitable effective models in addition to achieve a more transparent picture.

It is the purpose of this contribution to evaluate these quantities in the Nambu-Jona-Lasinio (NJL) model \cite{Klevansky1992,Vogl1991,Rehberg1996b} and to compare with similar approaches such as the Polyakov-Nambu-Jona-Lasinio (PNJL) \cite{Fukushima2003,Ratti2007} and the dynamical-quasiparticle model (DQPM) \cite{Peshier2004,Cassing2009a} as well as with the available lattice QCD data.

\section{Equation of state}

\begin{figure}
  \begin{center}
    \includegraphics[width=7.5cm]{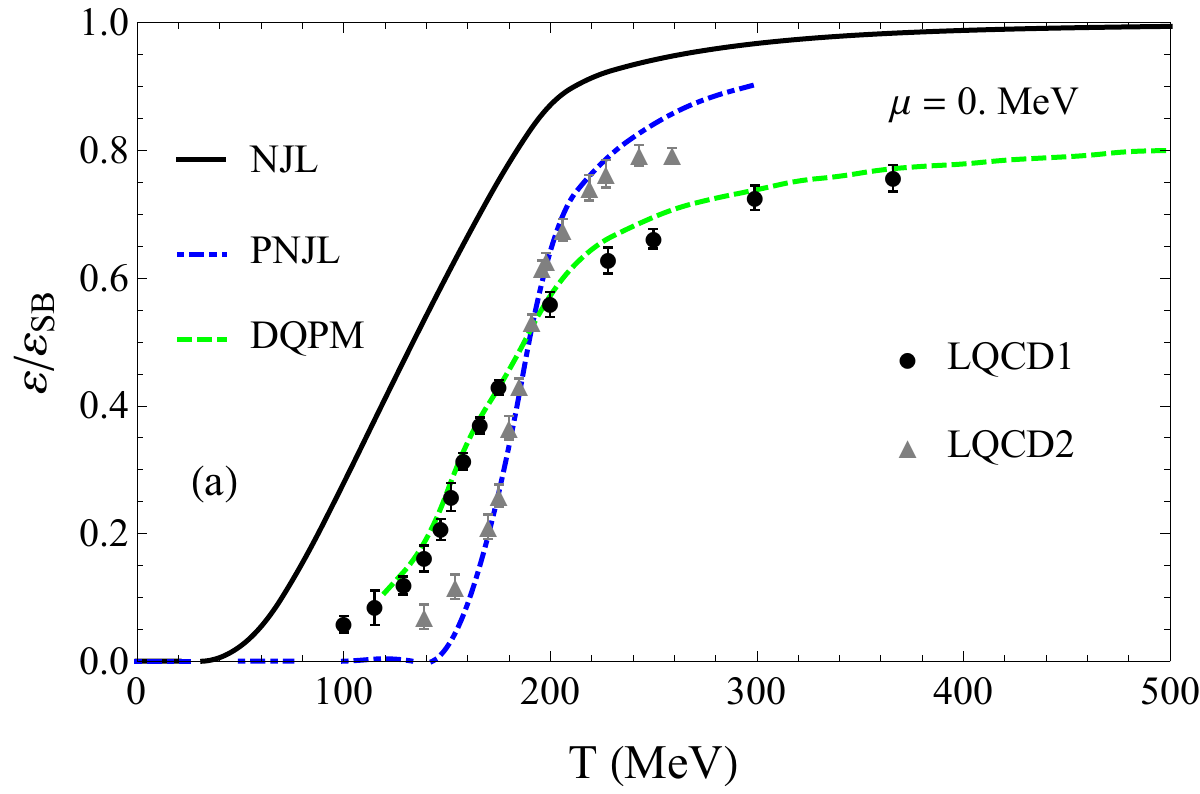} \ \ \
    \includegraphics[width=7.5cm]{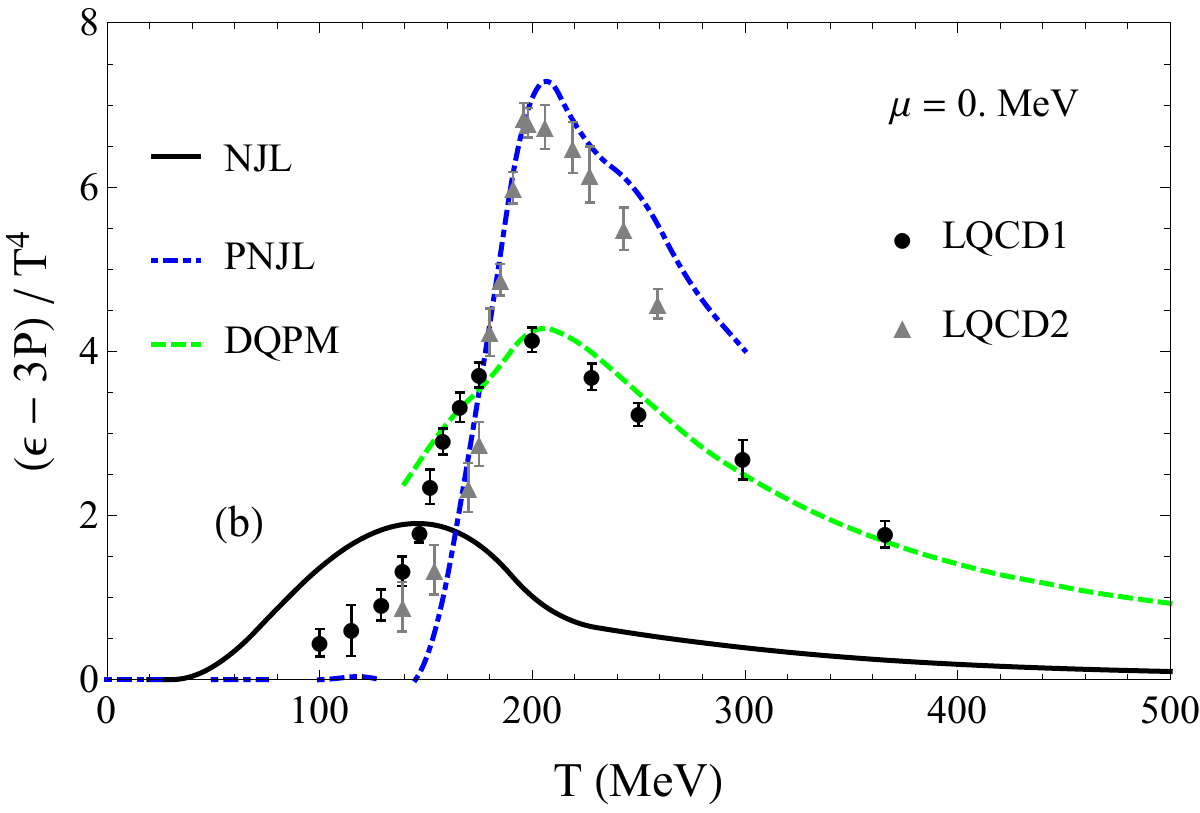}
  \end{center}
  \vskip -5mm
  \caption{Energy density $\varepsilon$ normalized to the energy density in the SB limit (a) and the trace anomaly $(\varepsilon - 3 P) / T^4$ (b) as a function of temperature $T$ from different models compared to the LQCD1 data from Ref. \cite{Borsanyi2010} and LQCD2 data from \cite{Cheng2010}. The PNJL results have been taken from Ref. \cite{Costa2010}.\label{EoS}}
  \vskip -2mm
\end{figure}

In order to compute the equation of state of strongly interacting matter, we use thermal distribution functions for partons (i.e. Fermi-Dirac and Bose-Einstein distributions). In the NJL model we compute the pressure $P$ and energy density $\epsilon$ (cf. Fig. \ref{EoS}) as a function of temperature $T$ using the relations based on the definition of the stress-energy tensor $T^{\mu\nu}$ for non-interacting particles at quark chemical potential $\mu$:
\begin{equation}
   T^{\mu\nu}(T,\mu) = g \int\limits_{0}^{\infty} \frac{d^3 p}{(2\pi)^3} f(E) \frac{p^\mu p^\nu}{E}.
\end{equation}

Note that in Fig. \ref{EoS}, which displays the resulting energy density $\varepsilon$ relative to the energy density in the Stefan Boltzmann (SB) limit, the SB limit is different for the NJL model and for PNJL/DQPM/LQCD due to the lack of the gluon degrees of freedom. We recall that the results of the Polyakov-NJL model differ from the NJL results due to the explicit gluon potential $\mathcal{U}$ \cite{Costa2010} which reflects the pressure from the gluons. The critical temperature $T_c$ is also different between the NJL, PNJL and LQCD since the parameters of the NJL are fixed at $T$=0.

In Fig. \ref{EoS}(b) we display the interaction measure -- known in LQCD as the trace anomaly -- in comparison with LQCD and find again the NJL model not to be in good agreement with the lattice data from Ref. \cite{Borsanyi2010,Cheng2010}. Note that the PNJL \cite{Costa2010} and DQPM calculations agree well with the LQCD data, however, both have been adjusted explicitly to different LQCD results, i.e. either to those from  Ref. \cite{Borsanyi2010} (DQPM) or from Ref. \cite{Cheng2010} (PNJL).

\section{Shear viscosity}

The shear viscosity $\eta$ is defined using the relaxation time $\tau$ in the dilute gas approximation for interacting particles as \cite{Chakraborty2010}:
\begin{equation}
  \eta(T,\mu) = \frac{1}{15 T} g_g           \int\limits_0^\infty \frac{d^3 p}{(2\pi)^3}
  \tau_g f_g \frac{{\bf p}^4}{E_g^2}
            +\frac{1}{15 T} \frac{g_q}{6} \int\limits_0^\infty \frac{d^3 p}{(2\pi)^3}
  \Bigg[ \sum_q^{u,d,s} \tau_q f_q + \sum_{\bar q}^{\bar u,\bar d,\bar s} \tau_q f_{\bar q} \Bigg] \frac{{\bf p}^4}{E_q^2}. \nonumber
\end{equation}

We find that the shear viscosity over entropy density ratio $\eta/s(T)$ in the NJL model shows a temperature dependence $\propto T^{-1}$ for high temperatures. Nevertheless, the order of magnitude is in agreement with the lattice QCD data from 1.2 $T_c$ up to 1.5 $T_c$. The $T^{-1}$ behavior of the viscosity in the NJL implies to go beyond the Kovtun-Son-Starinets (KSS) bound \cite{Kovtun2005}: $(\eta / s)_{\text{KSS}} = 1 / 4\pi$ above $T \sim 1.7 T_c$ which limits the applicability of the NJL model.

\section{Bulk viscosity}

The bulk viscosity defined in Ref. \cite{Chakraborty2010} reads in the relaxation time approximation (RTA)
\begin{equation}
  \begin{aligned}
  \zeta(T,\mu) =&\frac{1}{9 T} g_g           \int\limits_0^\infty \frac{d^3 p}{(2\pi)^3} \tau_g f_g
  \frac{1}{E_g^2}\Bigg[{\bf p}^2 - 3 c_s^2 \Bigg( E_g^2 - T^2 \frac{d m_g^2}{d T^2} \Bigg) \Bigg]^2\\
              &+\frac{1}{9 T} \frac{g_q}{6} \int\limits_0^\infty \frac{d^3 p}{(2\pi)^3}
  \Bigg[ \sum_q^{u,d,s} \tau_q f_q + \sum_{\bar q}^{\bar u,\bar d,\bar s} \tau_q f_{\bar q} \Bigg]
  \frac{1}{E_q^2}\Bigg[{\bf p}^2 - 3 c_s^2 \Bigg( E_q^2 - T^2 \frac{d m_q^2}{d T^2} \Bigg) \Bigg]^2.
  \end{aligned}
  \label{bulk}
\end{equation}

\begin{figure}
  \begin{center}
    \includegraphics[width=7.5cm]{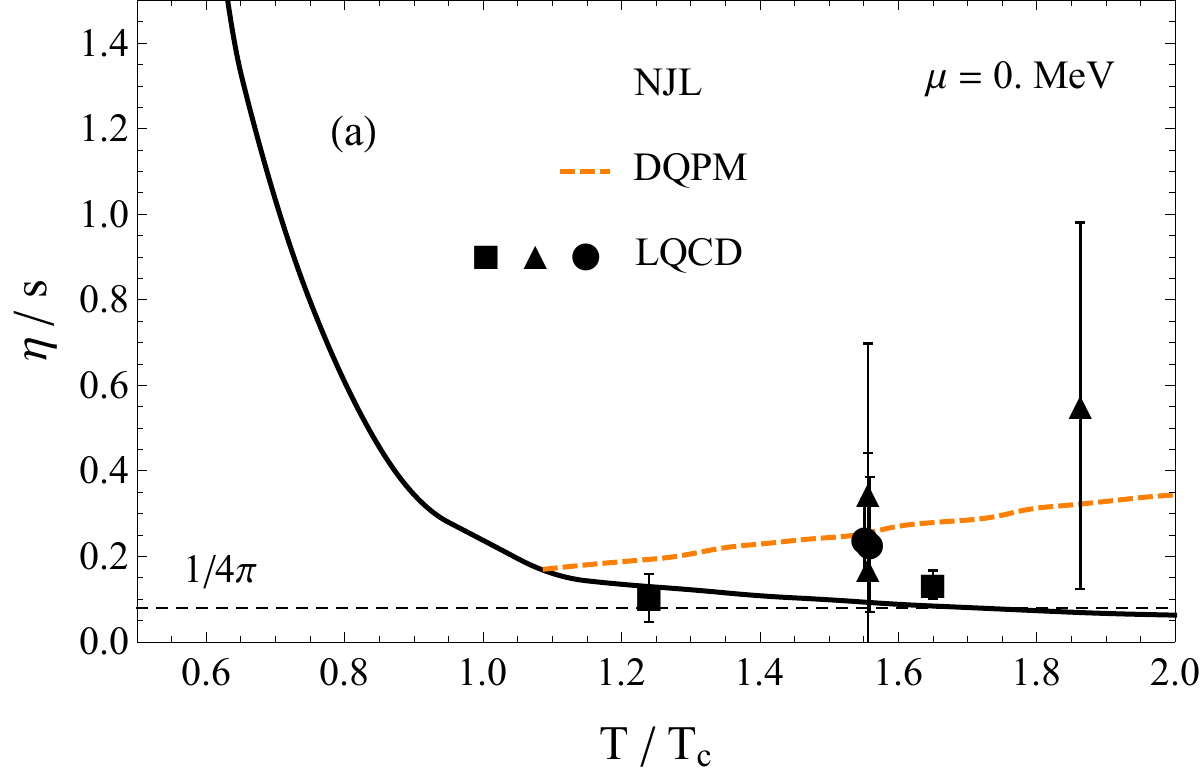} \ \ \
    \includegraphics[width=7.5cm]{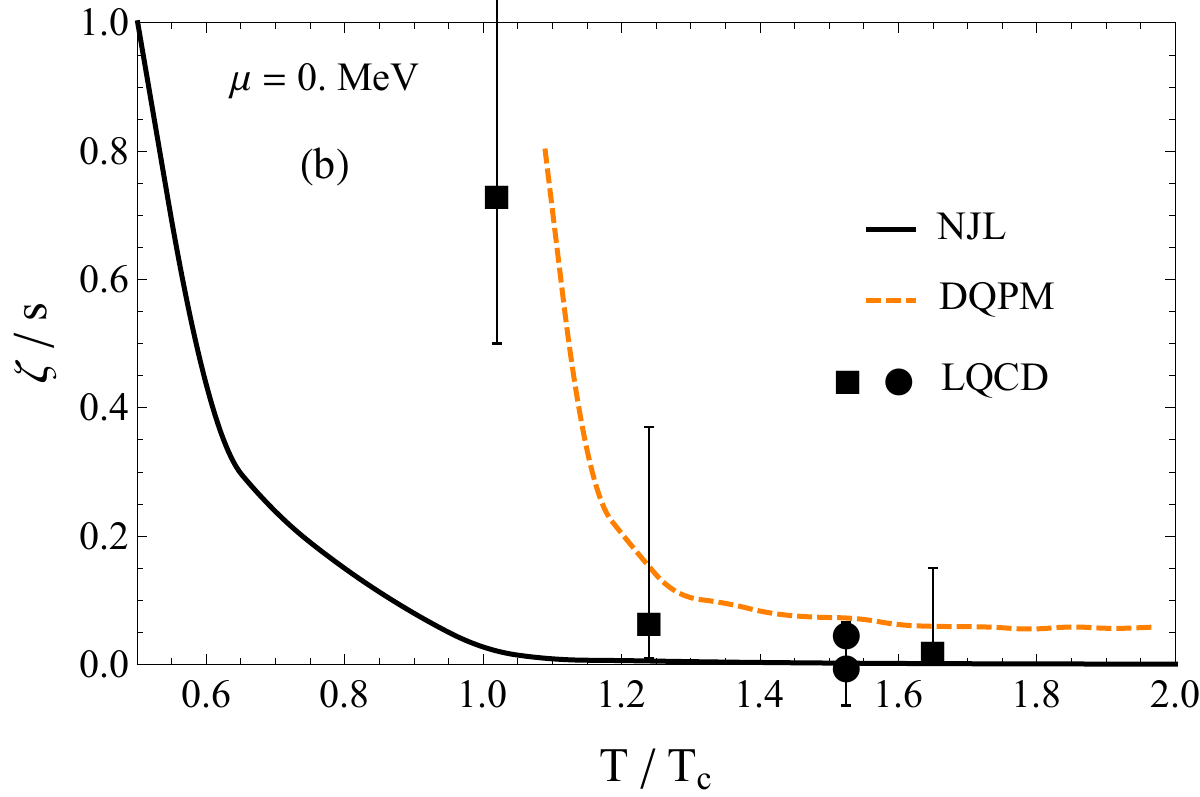}
  \end{center}
  \vskip -5mm
  \caption{Shear viscosity $\eta / s$ (a) as a function of $T / T_c$ compared to the LQCD data points from Ref. \cite{Meyer2007} (square), \cite{Nakamura2005} (triangle) and \cite{Sakai2007} (circle) and the result from the DQPM (dashed line); the bulk viscosity to entropy density ratio $\zeta / s$ (b) as a function of $T/T_c$ compared to the LQCD data points from Ref. \cite{Meyer2008} (square) and \cite{Sakai2007} (circle) and the DQPM (dashed line).\label{Viscosity}}
  \vskip -2mm
\end{figure}

The bulk viscosity over entropy density $\zeta/s$ from the NJL model is displayed in Fig. \ref{Viscosity}(b) and shows a very different temperature dependence  than $\eta/s$. Indeed, for high temperatures we find the limit $\zeta/s \to 0$. Moreover, the behavior around $T_c$ shows a peak in LQCD as well as in the DQPM. This peak is not seen in the NJL model (or shifted to much lower temperatures). This can be easily explained by the fact that the $T$-dependence of the masses of the degrees of freedom plays an important role for the bulk viscosity (\ref{bulk}).

\section{Electric conductivity}

The electric conductivity for charged particles  -- known as the Drude-Lorentz conductivity for a classical gas -- is defined as \cite{Reif1965,Cassing2013}:
\begin{equation}
  \sigma_e(T,\mu) = \sum_q \frac{e_q^2 \ n_q(T,\mu) \ \tau_q(T,\mu)}{m_q(T,\mu)},
  \qquad \text{with} \qquad
  e_q^2 = \frac{4\pi}{137} e^2,
\end{equation}
with $q = u,d,s,\bar u,\bar d,\bar s$, and $e = +2/3$ or $-1/3$ denoting the quark electric charge fractions.

Fig. \ref{Conductivity}(a) shows that for the DQPM as well as the NJL model the dimensionless ratio of the electric conductivity over $T$ is approximately linearly in $T$ for $T \geq T_c$ up to about 2 $T_c$ \cite{Cassing2013}. Both results are in a reasonable agreement with the present lattice QCD results although there is quite some uncertainty in the LQCD extrapolations.

\section{Heat conductivity}

The heat conductivity $\kappa$ is another quantity of interest that describes the heat flow in interacting systems \cite{Israel1979,deGroot1980} and only recently has regained interest in the context of relativistic heavy-ion collisions \cite{Denicol2012,Greif2013}.

The heat conductivity for charged particles is defined using the specific heat $c_V$ and the relaxation time \cite{Heiselberg1993}:
\begin{equation}
  \kappa(T,\mu) = \frac{1}{3} v_{\text{rel}} \ c_V(T,\mu) \ \sum_q \tau_q(T,\mu),
  \label{heatcond}
\end{equation}
with $q = u,d,s,\bar u,\bar d,\bar s$. For our purpose we assume $v_{\text{rel}} \simeq 1$ in the NJL model because the masses of quarks decrease with temperature $T$ whereas the mean momentum increases.

\begin{figure}
  \begin{center}
    \includegraphics[width=7.5cm]{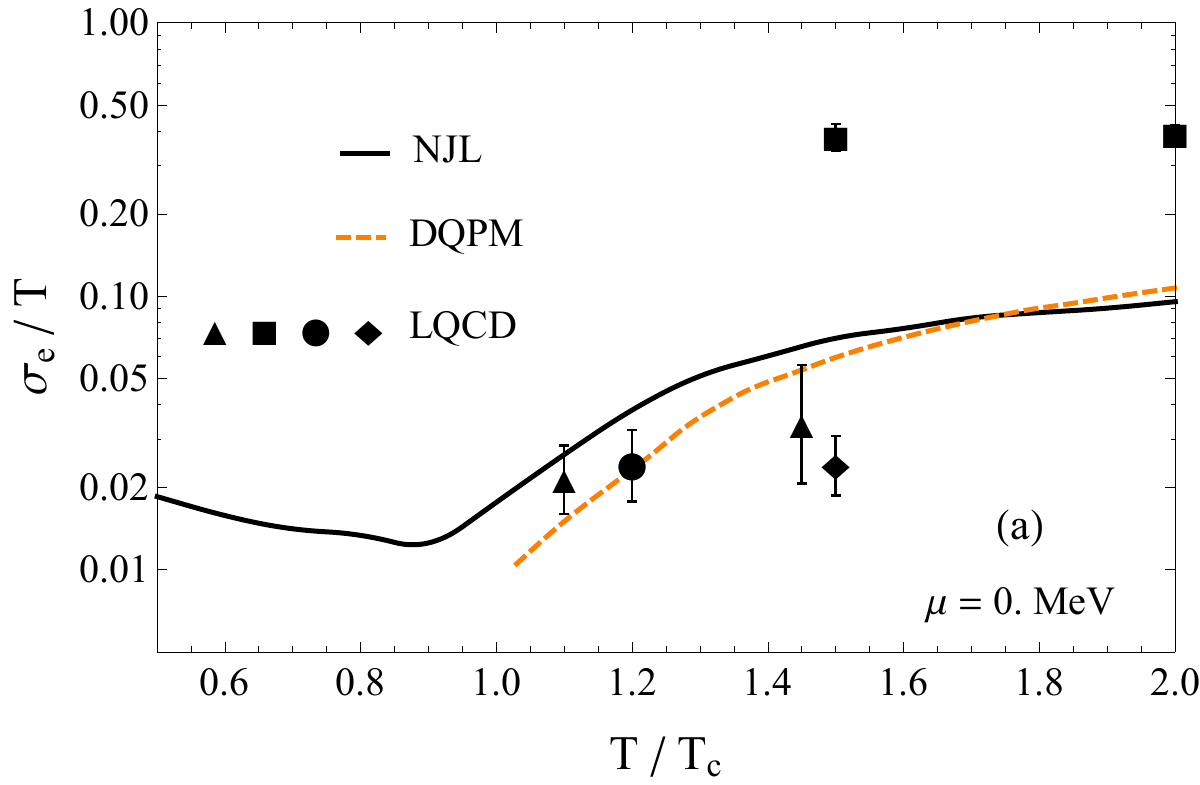} \ \ \
    \includegraphics[width=7.5cm]{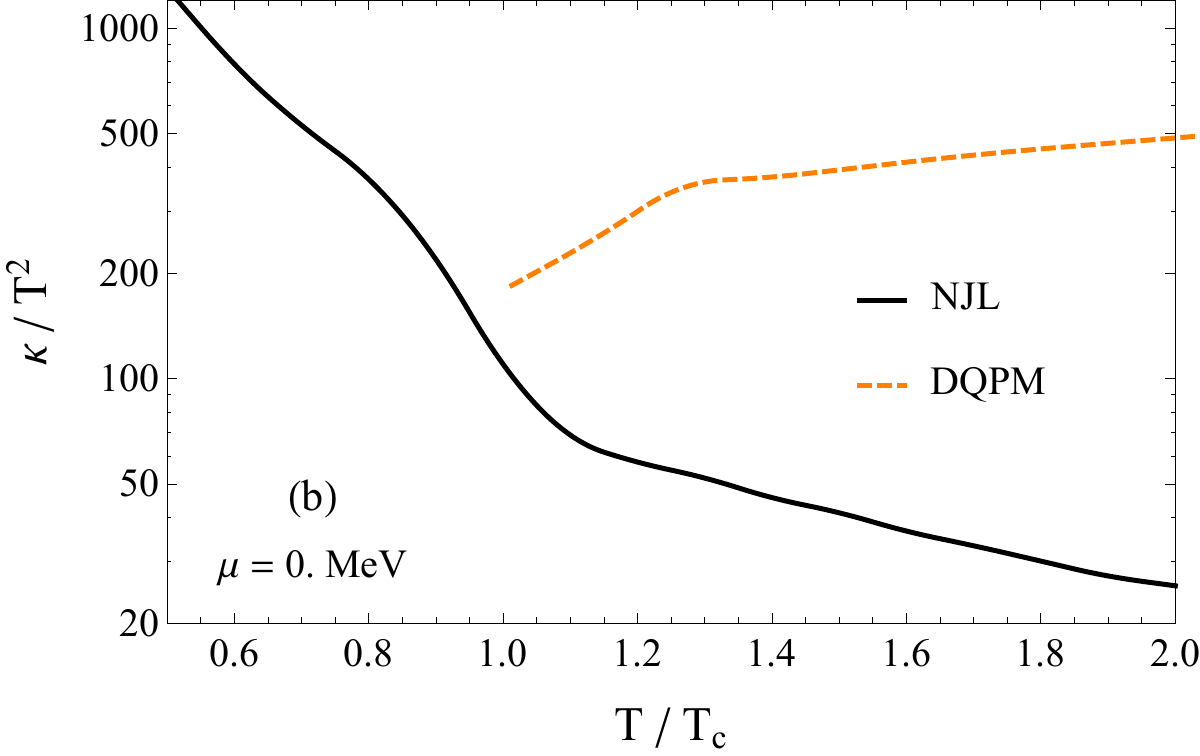}
  \end{center}
  \vskip -5mm
  \caption{The electric conductivity $\sigma_e / T$ (a) as a function of $T/T_c$ compared to the LQCD data points from Ref. \cite{Ding2013} (triangle), \cite{Aarts2007} (diamond), \cite{Gupta2003} (square) and \cite{Brandt2013} (circle) and the results from the DQPM (dashed lines); heat conductivity $\kappa / T^2$ (b) as a function of $T/T_c$ in the NJL model and the DQPM.\label{Conductivity}}
  \vskip -2mm
\end{figure}

Fig. \ref{Conductivity}(b) displays the dimensionless quantity $\kappa / T^2$ for both models. Whereas the DQPM shows a slightly rising ratio for $T_c < T < 2 T_c$ the NJL model predicts a rapid decrease with T for $T > T_c$. In this case no LQCD results are presently available.

\section{Conclusion}

In this study we have calculated thermodynamic properties of the NJL model for three flavors such as the energy density and pressure as a function of temperature $T$ up to a few times the critical temperature $T_c$. Furthermore, we have calculated the shear $\eta$ and bulk $\zeta$ viscosity as well as the electric conductivity $\sigma_e$ and heat conductivity $\kappa$ as a function of $T$ and compared to corresponding results from the DQPM, from the PNJL model and lattice QCD results when available.

\section*{References}

\bibliography{biblio}

\end{document}